\documentclass[3p,times]{elsarticle}
\usepackage{ifpdf}
\usepackage[usenames]{color}
\usepackage{graphicx}
\usepackage{amssymb}
\usepackage{amsmath}
\usepackage{hyperref}

\journal{Physica A}

\begin{document}

\begin{frontmatter}

\title{Identifying financial crises in real time}

\author[usp]{Eder Lucio da Fonseca}
\ead{eder.luciofonseca@usp.br}
\author[usp]{Fernando F. Ferreira}
\ead{ferfff@usp.br}
\author[ift,bdu]{Paulsamy Muruganandam}
\ead{anand@cnld.bdu.ac.in}
\author[ift]{Hilda A. Cerdeira\corref{hilda}}
\ead{cerdeira@ift.unesp.br}
\address[usp]{GRIFE-Escola de Arte, Ci\^{e}ncias e Humanidades, Universidade de
S\~ao Paulo, Av. Arlindo Bettio 1000,
03828-000 S\~ao Paulo, Brazil}
\address[ift]{Instituto de F\'{\i}sica Te\'orica, UNESP - Universidade Estadual
Paulista,
01140-070 S\~ao~Paulo, S\~ao Paulo, Brazil}
\address[bdu]{School of Physics, Bharathidasan University, Palkalaiperur Campus,
Tiruchirappalli
620024, Tamilnadu, India}
\cortext[hilda]{Correspondig author; Telephone: +55 11 3393-7852; Fax: +55 11
3393-7899}

\begin{abstract}
Following the thermodynamic formulation of
multifractal measure that was shown to be capable of detecting large
fluctuations at an early stage, here we propose a new index which permits us to
distinguish events
like
financial crisis in real time . We calculate the partition function from
where we obtain
thermodynamic quantities analogous to free energy  and specific heat. The
index is defined as the normalized energy variation and it can be used to study
the behavior of stochastic time series, such as financial market daily data.
Famous financial market crashes -
\emph{Black Thursday} (1929),  \emph{Black Monday} (1987) and
\emph{Subprime} crisis (2008) - are identified with clear and robust results.
The
method is also applied to the market
fluctuations of 2011. From these results it appears as if the apparent crisis
of
2011 is of a different nature from the other three. We also show that
the analysis has forecasting capabilities.

\end{abstract}

\begin{keyword}
Time series analysis, fluctuation phenomena, interdisciplinary physics,
econophysics, financial markets, fractals.

\PACS 05.45.Df,05.45.Tp, 89.65.Gh, 89.70.-a
\end{keyword}
\end{frontmatter}

\section{Introduction}

Events such as market crashes, earthquakes, epileptic seizures, material
breaking, etc., are ubiquitous in nature and society and generate anxiety and
panic difficult to control. Preventing them has been the dream for which many
scientists have devoted much of their work. There always exists the hope that
mathematics can supply tools and methods to predict when they will happen,
but it is necessary to understand the processes that produce
them.  It is well known that complex systems can generate
unpredictable outcomes while at the same time they present a series of general
features that can be studied. The financial market dynamics
belongs to this
class, therefore using these tools may help to solve the problem.

The usual dynamics of the financial market shows investors sending orders in
many different timescales, from high frequency
(minutes) to long-term (years) transactions. To describe it, one needs to take
an enormous number of unknown variables, although this does not impede us from
finding some temporal correlation or patterns in the time series. To address
these problems we shall use multifractal techniques known to capture non trivial
scale properties as shown by Mandelbrot (a pioneer to find self-similarity in
the cotton prices distribution~\cite{r01,r02}); by Evertsz \cite{r03}, who
confirmed the distributional self-similarity and suggested that market
self-organizes to produce such feature also; by Mantegna and Stanley \cite{r04},
who found a power law scaling behavior over three orders of magnitude in the
S\&P 500 index variation. These works were an important hallmark to shed light
in the financial market dynamics \cite{crepaldi}. It shows that we can use
concepts and tools
from statistical mechanics to model and analyze financial data
\cite{r05,grech,czar,preis1,preis2,ken1,ken2,pont}.  From there on many
researchers
are gathering
evidence that there exist alterations in the signal properties
preceding large fluctuations in the financial market
\cite{r06,r07,r08,r09,sorj,r20,sha}.
	
In this work we develop a new measure, the area variation rate (AVR), designed
with the goal of identifying financial crises in sufficient time to take
necessary steps. We do not attempt to explain the reasons of the crises,
which are outside our scope, but simply to introduce a systematic way to look at
the data which may help to distinguish systemic fluctuations -intrinsic to the
dynamics dictated by the internal interactions- from
those
generated by external inputs \cite{r11,r12}.

We handle time series in order to detect informational patterns in such a way
that it is not necessary to know the complete series, only past
data, in contrast with well established methods such as mutual information and
others \cite{r13,r14,r15}. This will make it particularly useful
for all practical predictive purposes. It will be shown that our method gives
some striking results when applied to financial market.

Our work is based in that of Canessa \cite{r16}, who adapted
calculations of time series onto previous results of Lee and Stanley \cite{r17}
on diffusion limited aggregation. In the latter, motivated by the analogy with
thermodynamics, a partition function is calculated, such that the elements are
the probability of a given event, which in our case will be
an increment in the time series. Applications of this method can be found in the
works by Kumar and Deo \cite{kumar}, Redelico and Proto \cite{rede} and Ivanova
et al \cite{r18}. This analogy
will allow us to call
some quantities by their corresponding thermodynamic variables, although keeping
in mind that it is just a convenience.

\section{Analysis of market data}

\begin{figure}[!ht]
\centering\includegraphics[width=0.49\columnwidth]{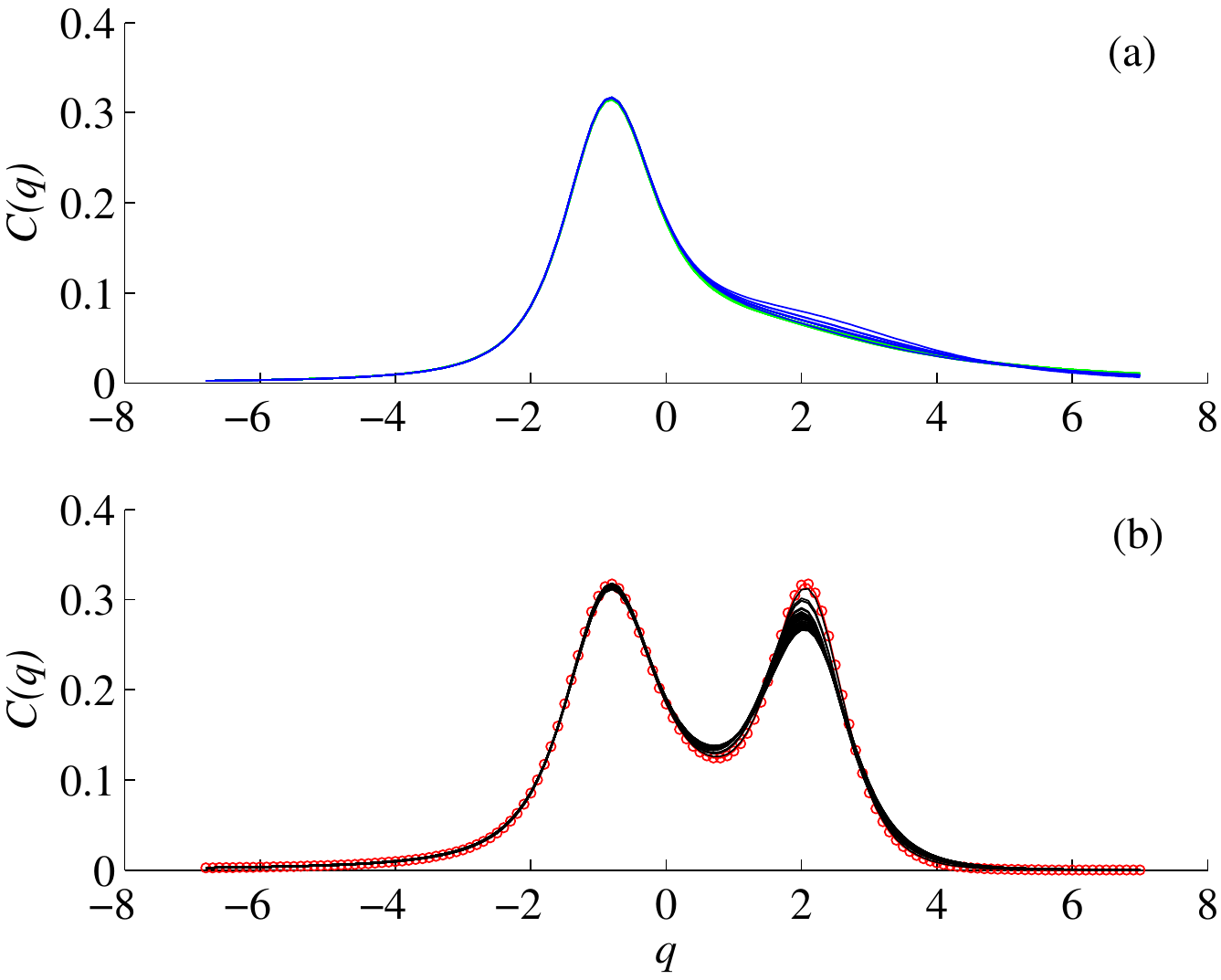}
\caption{Accumulated analogous Specific heat, $C(q)$, for the crisis of 1987,
Dow Jones index, (a) $50$ beginning with Black Monday (19 October 1987), (b)
$50$
days after \emph{Black Monday}. Dots correspond to  $BM$. Here $N=1000$, $T=1$
and $l=1$.}
\label{fig1}
\end{figure}

In order to analyze the data we proceed as follows:

Given a series of $N$ elements $x(t)$, we determine the increments $\Delta
x(t,T) = x(t+T) - x(T)$, where $t=1,2,\ldots,N$ and $T$ is an adjustable
quantity. We define a measure \cite{r16,r17}
\begin{align}
\mu_t = \frac{\left \vert \Delta x(t,T) \right \vert}{\sum_{t=1}^{N} \left \vert
\Delta x(t,T) \right \vert},
\end{align}
and a generating function (partition function) as a sum of the $q-th$ order
moments of this measure \cite{r17}
\begin{align}
Z(q,N) \sim \sum_{t=1}^{N} \mu_t^q \sim N^{-\tau(q)},
\end{align}
where
\begin{align}
\tau(q) = (q-1)D_q,
\end{align}
$\tau(q)$ plays the role of a free energy, and $D_q$ is the fractal
dimension of the system. $\tau(q)$ is related to the generalized Hurst exponent
\cite{kumar,bar} and to the R\'enyi entropy \cite{renyi}. Following this path,
Canessa \cite{r16}, searching to
model economic crises with nonlinear Langevin equations, showed that for
financial market data the quantity
\begin{align}
C_q = - \frac{\partial^2 \tau(q)}{\partial q^2},
\end{align}
presents special effects when the string of economic data includes that of a
crash.  We shall refer to this quantity as \emph{analogous specific heat}
(ASpH). Using the S\&P500 data corresponding to the 1987 market crash known as
\emph{Black Monday} (BM), it has been shown in Ref. [26] that these curves
present two
lobes when the time series of $N$ elements contains the data point of the
crash. 
Further,  the maxima of ASpH, $C(q)$, diminishes when the single data point from
the day of the crash is removed and it disappears when the data of the
neighboring days is deleted \cite{r16}. This lobe is due to the presence of
large
fluctuations ($q>0$) in the portion of the time series calculated. In
Fig.~\ref{fig1} we show accumulated curves for $C(q)$ for Dow Jones data for
windows of time before and after the crash, in particular Fig.~\ref{fig1}(a)
represents the accumulated curves for 50 days before the crash and
Fig.~\ref{fig1}(b), for 50 days after -and including- the
crash. The dotted
curve corresponds to BM day. Further in the text we will complete the
description of these curves.	
	
\begin{figure}[!ht]
\centering\includegraphics[width=0.49\columnwidth]{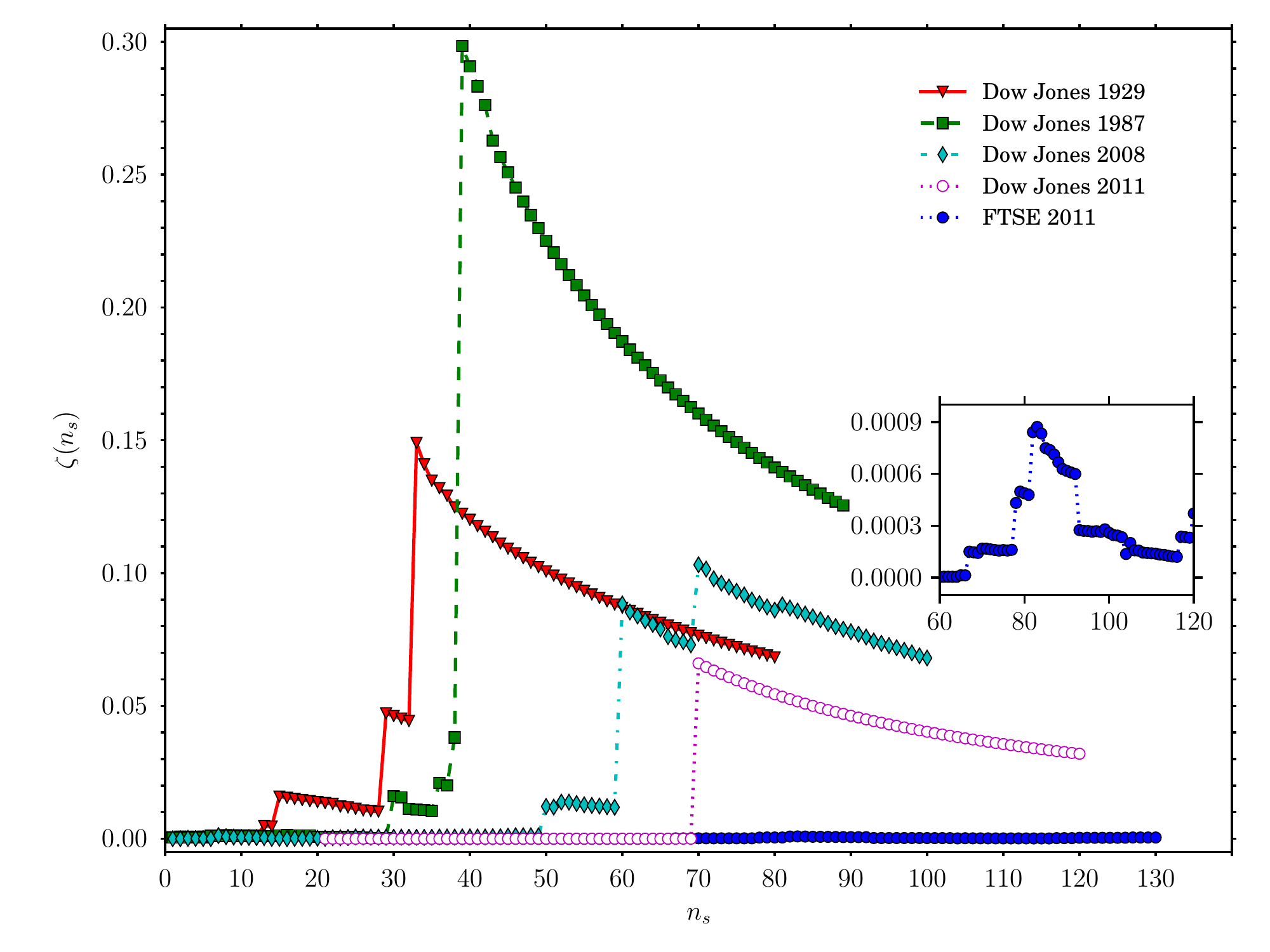}
\caption{Plot of the area variation rate  $\zeta(n)$ for the daily index for
different events: Crash of '29, $n_s=30$ corresponds to 24 October 1929 (red
triangles) and $n_O=-20$; crisis of '87, $n_s=40$ refers to 19 October 1987 (green squares), $n_O=-10$;
crisis of $2008$, $n_s=50$ represents the bankrupcy of Lehman Brothers (cyan
diamonds) and $n_O=0$; events of
2011, we show the AVR of  the  Dow Jones data where $n_s=70$ corresponds to 5
August
2011 (lilac empty circles) and $n_O=20$ and that of the FTSE100 series, which is non
noticeable in this graph (blue filled circles), $n_O=30$. For this case we have used
$T=1$, $l=1$ and $N=1000$.}
\label{fig2}
\end{figure}

To study the differences in the ASpH along the complete time series of size $L$,
let us consider a window of size $N$ as a section of the series
to be analyzed.  This window is shifted along the data, with a given shift $l$,
less or equal to $N$. $T$, $N$ and $l$ are parameters to be chosen for the
particular system under
consideration and the values are not predetermined.  They are chosen as the best
set for a given type of data, their selection will depend on the
stability of results which will be discussed below. We
calculate the area
under the curve of ASpH, which we call $A(n)$, where $n$ is the index that
identifies the window where our calculation is done, that is the number of
shifts that we had performed. To identify the corresponding index in the
complete series we notice that
\begin{align}
t'_n = t_0 + N + n \, l,
\end{align}
where $t_0$ is the time where the calculation starts, and $n=1$ corresponds
to $t'_1=t_0+N$ of the series considered. Since the areas are
calculated over a long period of time they possess a long memory, which is not
necessarily desirable when we are near a crash or in a succession of them. In
order to obtain seemingly uniform information we define the area variation rate
(AVR) $\zeta(n)$ as
\begin{align}
\zeta(n) = \left\vert \frac{A(n)}{\bar A} - 1 \right\vert,
\end{align}
where the mean $\bar A$ is calculated over all the windows previous to $n$.
When
there is no particular event in a time series this
parameter fluctuates around or below
$10^{-3}$ - see inset in Fig.~\ref{fig2}. Although we do not have
any proof, this seems to be always the case and it agrees with the AVR generated
by white noise as will be seen below.

To apply this method we use daily closing indexes from
different markets: Dow Jones, DAX, FTSE100,
CAC, S\&P 500 and Nasdaq, although not all results are shown. The data was
obtained from finance.yahoo.com. The complete URL's and the dates of downloading
the data are given
in reference \cite{r19}.
The AVR for financial market daily data over a period of $100$ days is drawn in
Fig.~\ref{fig2} for several well
known events: the crisis of 1929, 1987, 2008 and 2011. All plots start $50$ days
before the event that
history identifies as the day of the corresponding crash, as done for
Fig.~\ref{fig1}.  Thus, all the data
used for the
calculation of the AVR of a given day is prior to the date. In
order to be able to notice their common features as well as their differences
all the curves are plotted in the same graph with shifted origins$(n_O)$. The crash of '29 is shown in
red triangles, where $n_s=n+n_O=30$ corresponds to 24 October 1929, known as \emph{Black
Thursday} (BT).
The
curve in green squares
plots the same
data, $n_s=40$ marks 19 October 1987, or \emph{Black Monday} (BM).  Next, in
cyan diamonds, we present the same parameter AVR
for the financial crisis of the year 2008, which was called the \emph{Subprime
Crisis}, and compared by the Press to the crash of '29.  Although
there is a jump in the curve, it is well below the other
curves.  The value of $n_s=50$ corresponds to  15 September 2008, when Lehman
Brothers announced bankruptcy, event
which may have produced the succesive jumps at $n_s=60$ and $n_s=70$. Next, in lilac
empty circles is the
area variation rate for the Dow Jones 2011 data, where $n_s=70$ corresponds to 5
August. This date was chosen since on that day Standard and Poor's announced
the decrease in the credit ratings of the United States \cite{stdp}.
Therefore it is only
natural to expect some reaction of the market.
 This is the only event
worth mentioning in
the Dow Jones data during 2011. Finally,
in blue
filled circles we show FTSE100 during 2011, which stays below $10^{-3}$
throughout the year (partially shown in the inset of Fig.~\ref{fig2}). Apart
from Dow Jones and FTSE100, we analyzed other European markets for 2011: DAX
(Germany) and CAC (France). All of them, except the FTSE100,
have some fluctuation above average in
the summer of 2011 and a noticeable jump around 6 December 2011 - day when the
results of a discussion on
the survival of
the Euro was announced. All fluctuations are within the
same order of magnitude but below 1.5\%.

\begin{figure}[!ht]
\centering\includegraphics[width=0.75\columnwidth]{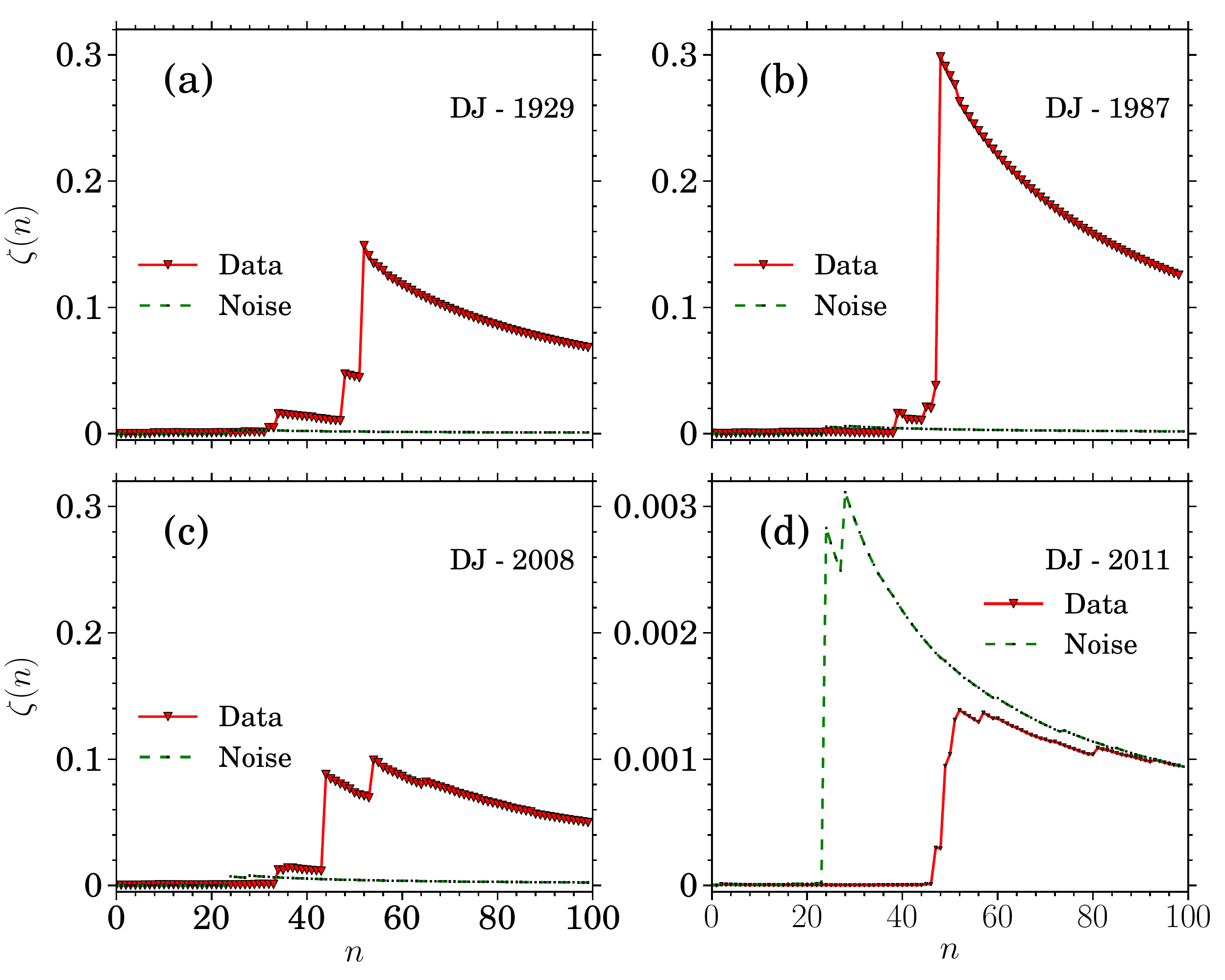}
\caption{Plot of the area variation rate $\zeta(n)$ for the Dow Jones daily
index and for randomly generated time series with the same average and variance
of
the corresponding Dow Jones index with parameter values  $l=1$, $T=1$, $N=1000$
for: a) Crash of 1929; b) Crash of 1987; c)
Subprime crisis 2008; d) Data centered around 5 August 2011 }
\label{fig3}
\end{figure}

From that figure we notice that the AVR rises above noise at least one week
before the crash for 1929, 1987 and 2008, characteristic that may be used
as an indicator of the coming event. There is no significant fluctuation in the
AVR during 2011. To distinguish if the increase is related to the crash we
compare these results with the AVR index of a randomly generated time series
(noise).
The data used to calculate the AVR  of the noise was obtained from a long
series generated from white noise (independent and normally distributed), with
the same mean and variance of the corresponding Dow Jones index. A value that
``rises
above noise'' is at least one order of magnitude larger than these fluctuations.
Results for the four cases discussed previously can be seen in
Fig~\ref{fig3}. It is interesting to notice that noise and 2011 fluctuations are
of the same order of magnitude.

\begin{figure}[!ht]
\centering\includegraphics[width=0.75\columnwidth]{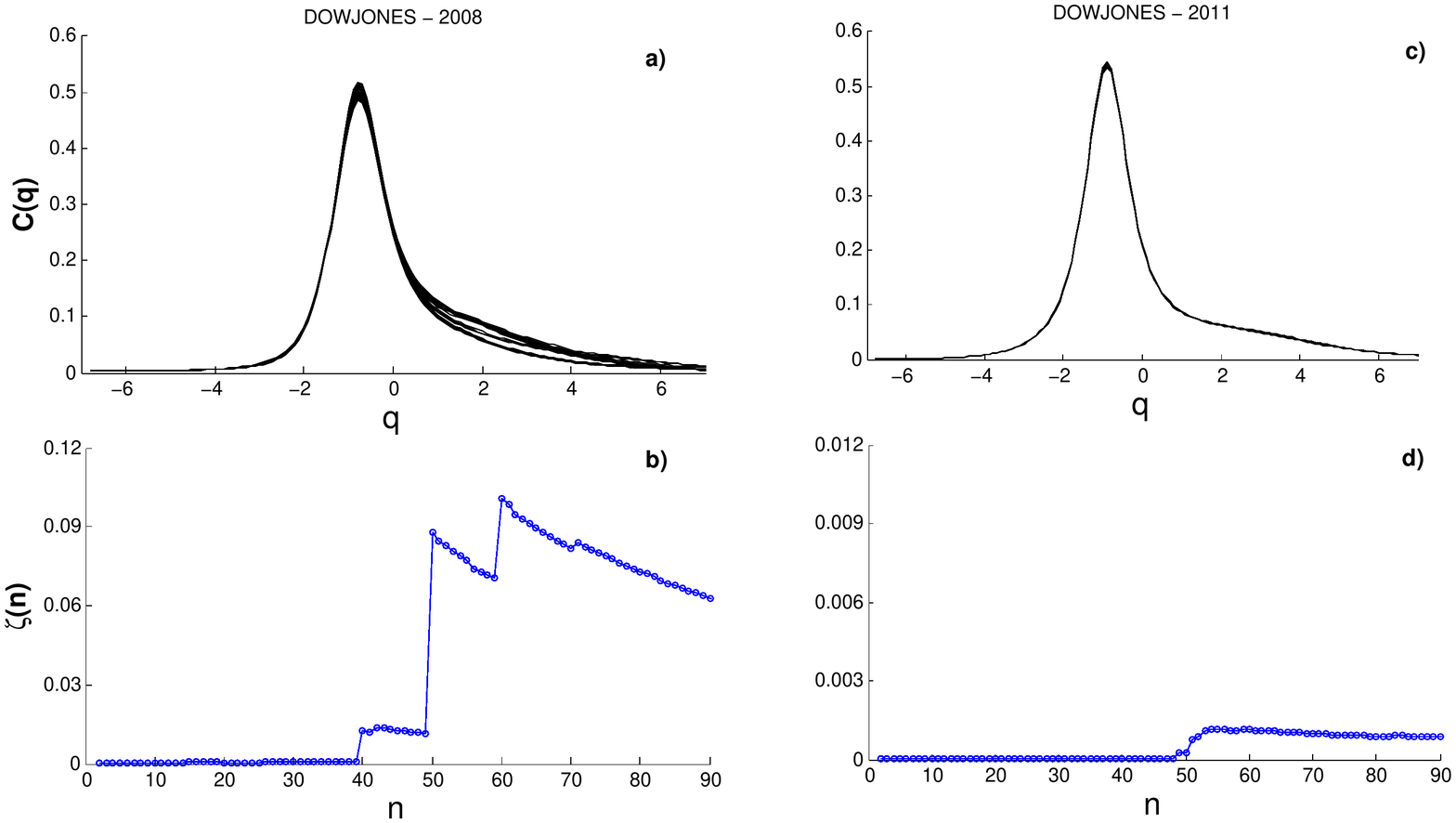}
\caption{(a) Analogous specific heat, $C(q)$, for Dow Jones data corresponding
to the Subprime crisis, 2008. (b) Plot of the corresponding area variation rate,
$\zeta(n)$. (c) $C(q)$ for Dow Jones 2011. (d) Plot of the area variation rate,
$\zeta(n)$, for 2011, the jump corresponds to 5 August 2011.All curves were
calculated with  $l=1$, $T=1$, $N=1000$}
\label{fig4}
\end{figure}

To interpret the importance of the AVR in the identification of crises, we
recall that in Figs.~\ref{fig1}(a) and \ref{fig1}(b) we show the accumulated
curves for the ASpH individually calculated over each window of length $N$
previous to the point $n$.  When we move to $n+1$, we slide the window
onwards a
quantity $l$ (shift) and calculate $C(q)$ again. There are $50$ overlapping
curves in each figure and they are both types: single and
doubled lobes. There is a clear change of shape in the curve
corresponding to BM, already identified by Canessa by exclusion of the crash
day's data in his calculation. The distinct shape that the curve has on the 19th
October shows the sensitivity of the method to the inclusion of the data of the
crash since it is only one point within $N$ that produces a dramatic change
in the figures. But the most important feature in the identification of the
crisis is that the second peak of the curve $C(q)$, for $q>0$, appears at the
same moment when the curve AVR jumps, and it does only if the jump is large,
thus clearly pointing at the crash, which is produced by the very large
fluctuations. Therefore large fluctuations in the market data give rise to the
second maximum in the ASpH, which originate a discontinuity in the AVR, while
robustness in $N$, $T$ and $l$ help to identify a crisis.

Next, we analyze the accumulated curves $C(q)$ for all the
cases shown in
Fig.~\ref{fig2}, we have found a clear second lobe for the crises of '29 and
'87. In particular let us focus our attention in the
accumulated ASpH calculated along $n$ for 2008 and 2011 (Fig. ~\ref{fig4}(a) and
Fig.~\ref{fig4}(c), respectively) and the corresponding area variation rate as a
function of $n$ - Fig.~\ref{fig4}(b) and  Fig.~\ref{fig4}(d), respectively.
There
is no well defined second maximum in ASpH for 2008, although
the results are robust against changes in $N$ and $T$ (not shown). For 2011 we
detect no significant jumps in the AVR, no changes in the ASpH and the results
do not persist under changes in the parameters (shown next).

\begin{figure}[!ht]
\centering\includegraphics[width=0.75\columnwidth]{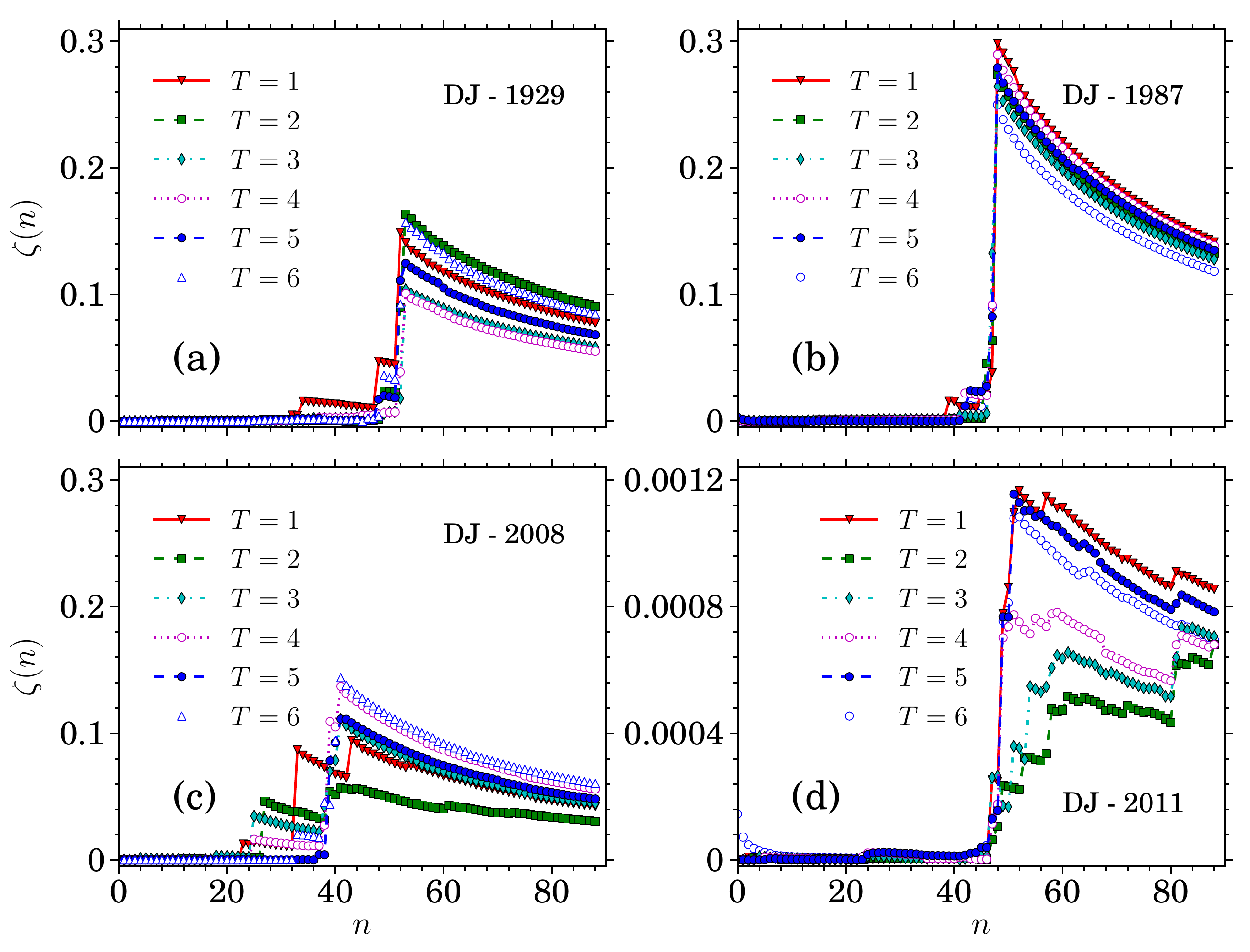}
\caption{Area variation rate, $\zeta(n)$, as a function of $n$, for different
values of $T$: (a) Dow Jones crash October 1929; (b) Dow Jones crash October
1987; (c) Dow Jones
Subprime crisis of 2008; (d) Dow Jones August 2011. All curves were calculated
with  $l=1$ and $N=1000$}
\label{fig_T}
\end{figure}

\begin{figure}[!ht]
\centering\includegraphics[width=0.75\columnwidth]{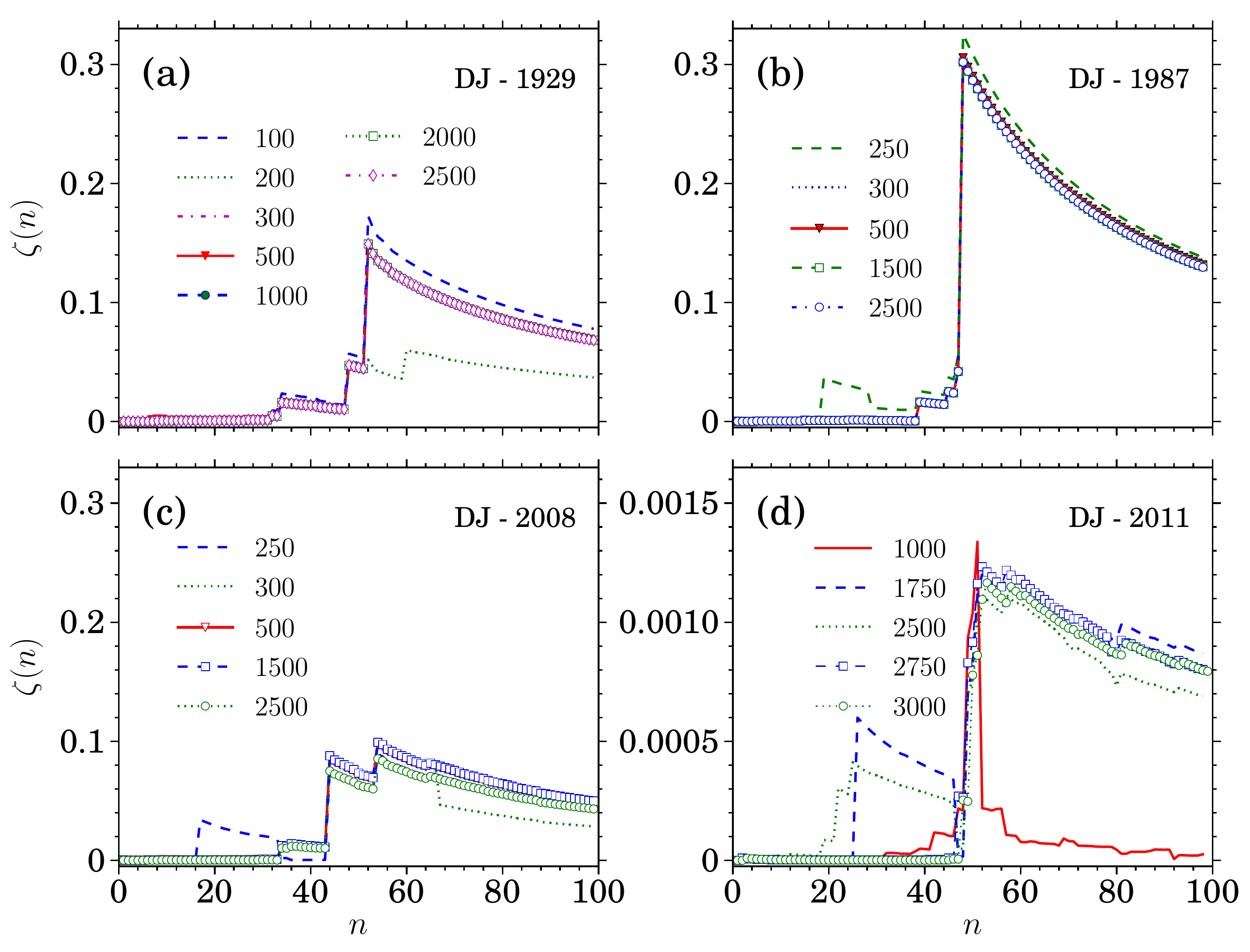}
\caption{Area variation rate, $\zeta(n)$, as a function of $n$, for different
values of $N$: (a) Dow Jones crash of 1929; (b) Dow Jones of 1987; (c) Dow Jones
2008; (d) Dow Jones August 2011. All curves calculated with  $l=1$ and $T=1$}
\label{fig6}
\end{figure}

A small size window may produce false positives. Investigating the
Dow Jones data over a century, it is noted that sometimes the
AVR presents
unexpected jumps. However, most of those jumps do not exhibit the pattern
observed in the famous crashes and the jumps tend to vanish increasing the
window size $N$ and $T$. In Fig.~\ref{fig_T} results are shown for the AVR of
all cases when varying the time lag. Results for 1929, 1987 and
2008 remain stable. For 2011, the jump on 5 August changes shape while staying
always below or around noise level. We
show similar results but varying N in Fig.~\ref{fig6}, starting at a very low
window size, $N=100$ for 1929. The AVR for 1929, 1987 and 2008 rapidly converges
to a final value. That for 2011 oscillates without ever reaching a stable value.
 Another important
factor is that the features previous to the crash discussed before appear for
all values of $N$ and $T$ with $T=1$ giving the best results. Most features
disappear when the shift $l$ becomes of the order of $N$, which is expected.
From these it appears as if the crisis of 2011 is of a different nature from the
other three. Calculations for different parameters $N$ and $T$ will show a real
crisis, real being defined as stable against variations of $N$, $T$ and $l$ by
comparison with the events historically recognized as crises.

\section{Summary}

Summarizing, we have presented a method by which market crisis
can be
identified and distinguished from simple scares which could avoid diffusion of
panic. We noticed that when a crisis is systemic
it behaves as in the cases well recognized historically as crises. It is usually
accompanied by a sizable increase in the area variation rate by several orders
of magnitude. It corresponds to a crash only if the analogous specific heat
also presents a second maximum. Finally, all results are stable under changes in
$N$,
$l$ and $T$. We notice that for some well known crisis such as the crash of '29
and '87, the AVR already shows a substantial increase above noise almost two
weeks before the event creating expectations as to the possibility of being used
as a warning signal. While it is quite clear that the existence of large
fluctuations ($q>0$) produces stable jumps in AVR, the predictive ability of the
algorithm seems to depend on the values of $T$. An important factor of this
method is that all calculations are done using data preceding the singular
event, thus making it extremely useful, in particular for medical cases such as
epilepsy. Its identification through a electroencephalogram, performed by the
recognition of a specific characteristic of the temporal series, is only done
after the episode. A method based on the calculation of the AVR may help in
predicting seizures.
 Although this is not a tool which can help to avoid crises, due to the
simplicity of its use, it can be implemented to monitor economic or
physiological data
under the possibility of a crisis. More research is underway to study
applications -- in particular the possibility of foreseeing epileptic seizures
--
and to establish
the algorithm in a sound mathematical basis.

\section*{Acknowledgments}

HAC thanks E. Canessa for calling her attention to his work. The authors thank
Roberto Kraenkel and A. Christian Silva for critical reading of the manuscript.
ELF acknowledges support from CAPES (Coordena\c{c}\~{a}o de Aperfei\c{c}oamento
de Pessoal de N\'{\i}vel Superior).
PM thanks the Theoretical Institute of Physics (IFT-UNESP) for
hospitality.
The work of PM is supported by DST and CSIR (Government of India) sponsored
research projects.

\end{document}